\documentclass[superscriptaddress,prl,reprint]{revtex4-1}
\usepackage{graphicx}
\usepackage{float}
\usepackage{array,url}
\usepackage{units}
\usepackage{epstopdf}
\usepackage{setspace}
  \usepackage{braket}
  \usepackage{amsmath}
\usepackage[usenames,dvipsnames]{color}
\usepackage{soul,xcolor}

\newcommand{\bracket}[2]{\ensuremath{\langle {#1} | {#2} \rangle}}

\newcommand{\aver}[1]{\ensuremath{\langle {#1} \rangle}}

\begin{document}

\title{
Heralded Interaction Control between Quantum Systems}

\author{Yiheng Duan}
\affiliation{Department of Physics and Research Laboratory of Electronics, Massachusetts Institute of Technology, Cambridge, Massachusetts 02139, USA}

\author{Mahdi Hosseini}
\affiliation{Birck Nanotechnology Center, School of Electrical and Computer Engineering, Purdue University, West Lafayette, Indiana 47907, USA}

\author{Kristin M. Beck}
\affiliation{IonQ, College Park, Maryland 20740, USA}

\author{Vladan Vuleti\'c}
\email{vuletic@mit.edu}
\affiliation{Department of Physics and Research Laboratory of Electronics, Massachusetts Institute of Technology, Cambridge, Massachusetts 02139, USA}

\date{\today}
\begin{abstract}
Quantum mechanical expectation values for subsets can differ substantially from those for the whole ensemble. This implies that the effect of interactions between two systems can be altered substantially by conditioning. Here we experimentally demonstrate that, for two light fields $\psi_S$ (signal) and $\psi_A$ (ancilla) that have only weakly interacted with one another, subsequent measurements on the ancilla can produce substantial conditional amplification, attenuation, or phase shift of $\psi_S$. We observe conditional signal power changes within a factor of 30, and phase shift up to $\pi/2$, induced by small changes in the ancilla measurement basis. The method is generically applicable to a variety of systems, and allows one to modify or boost a given interaction by trading in success probability for interaction strength. 
\end{abstract}

\maketitle

In quantum mechanics, rare measurement outcomes can have surprising consequences~\cite{Spin100}. Here we consider two quantum systems, $S$ (signal) and $A$ (ancilla), that are made to interact weakly, as characterized by an interaction-induced moderate average change $\langle\delta s\rangle$ of some quantity $s$ associated  with the signal system $S$. Now assume that there is some binary measurement basis for the ancilla system $A$, such that outcome $a_1$ with probability $p_1 \ll 1$ is observed rarely compared to outcome $a_0$ with probability $p_0 \approx 1$. If we also assume that $a_0$ is associated with no change in the system parameter $s$ ($\delta s \vert_{a_0}=0$), then in those rare occasions when outcome $a_1$ is observed for the ancilla system, there must be an associated very large signal change $\delta s \vert_{a_1} \propto 1/p_1$ to reproduce the average change $\aver{\delta s}=p_1 \delta s \vert_{a_1}$ when the ancilla system is not measured. A different measurement basis of the ancilla system can then give rise to different conditional changes in $s$, or induce large changes in an altogether different system parameter $s'$. Thus one can  think of the measurement basis of $A$ as conditionally controlling the type and strength of the interaction between $S$ and $A$. Thus, at the expense of success probability, one can modify the quantum state of the signal system $S$ and its observables far beyond the changes induced by the average (unconditional) interaction, and one can choose which observables are conditionally controlled.
  
\begin{figure}[!th]
\centerline{\includegraphics[width=0.95\columnwidth]{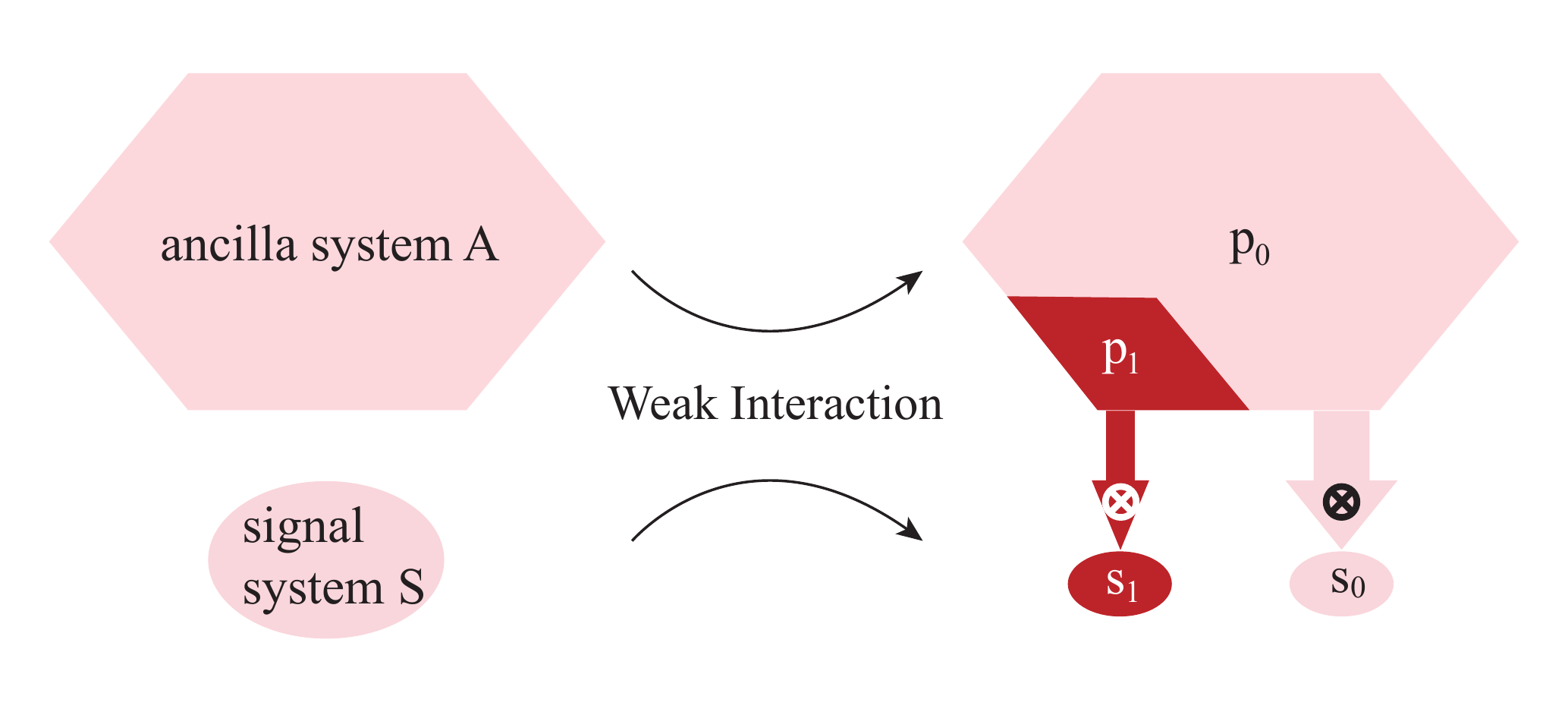}}
\caption{Boosting a weak interaction conditionally. A weak interaction between a signal system $S$ and an ancilla system $A$ results in a small average shift $\langle\delta s\rangle$ of some signal quantity $s$. Assume that for binary measurement outcomes $a_0$, $a_1$ of $A$, if $a_0$ is observed with probability $p_0 \approx 1$, the system parameter $s$ will maintain its value before the interaction. If, on the other hand, $a_1$ is measured (probability $p_1 \ll 1$), the associated signal $s=s_1$ can be much different from $s_0$, and the conditionally prepared state of $S$ can differ substantially from the input state: at the expense of success probability, a strong heralded interaction is realized.}
\label{PIC}
\end{figure}
  
Such heralded interaction control (HIC) can be viewed as an extension and generalization of weak-measurement \cite{Pryde2005,hosten2008spinhall,xu2013phase,merano2014observing,magana2014amplification} and noiseless-amplification schemes \cite{Ferreyrol2010,usuga2010,zavatta2011,osorio2012,kocsis2013,2016expnoiseamp}, and can be used for a variety of purposes in quantum engineering. Noiseless amplification of coherent optical states \cite{usuga2010,Ferreyrol2010,zavatta2011,osorio2012,kocsis2013,2016expnoiseamp} can be viewed as HIC. By coupling light fields to other systems, HIC allows one to magnify and measure tiny physical quantities in the presence of technical noise (weak-measurement schemes) \cite{Pryde2005,hosten2008spinhall,xu2013phase,merano2014observing,magana2014amplification}. When applied to large systems such as the collective spin of an atomic ensemble, even a single photon can be used to control the atomic spin, and conditionally prepare it in a desired collective entangled spin state ~\cite{mcconnell2015entanglement,carve,frowis2017}.
  
In this Letter, we report how a weak optical nonlinearity can be conditionally boosted to affect large amplitude or phase changes of a (weak) signal light field. We first weakly entangle two optical modes in a cavity quantum electrodynamics setup, and use HIC to coherently amplify, attenuate, and change   the phase of the signal mode within a large parameter space. We modify the average photon number $\aver{n_s}$ in the signal mode over a range of 30, ($\aver{n_s}$ changed by factor between 0.1 to 3.2), and the phase between 0 and $\pi/2$. These conditional changes of the signal mode are accomplished under conditions of weak interaction with the ancilla mode, where the average unconditional photon number and phase change are as small as $\aver{\delta n_s}=-1.3$\% and $\aver{\delta \phi_s}=\pi/80$, respectively. We further show that a small change in the (polarization) measurement basis of the ancilla mode by a few degrees can produce a large change in the signal state.

\begin{figure}[!th]
\centerline{\includegraphics[width=1.\columnwidth]{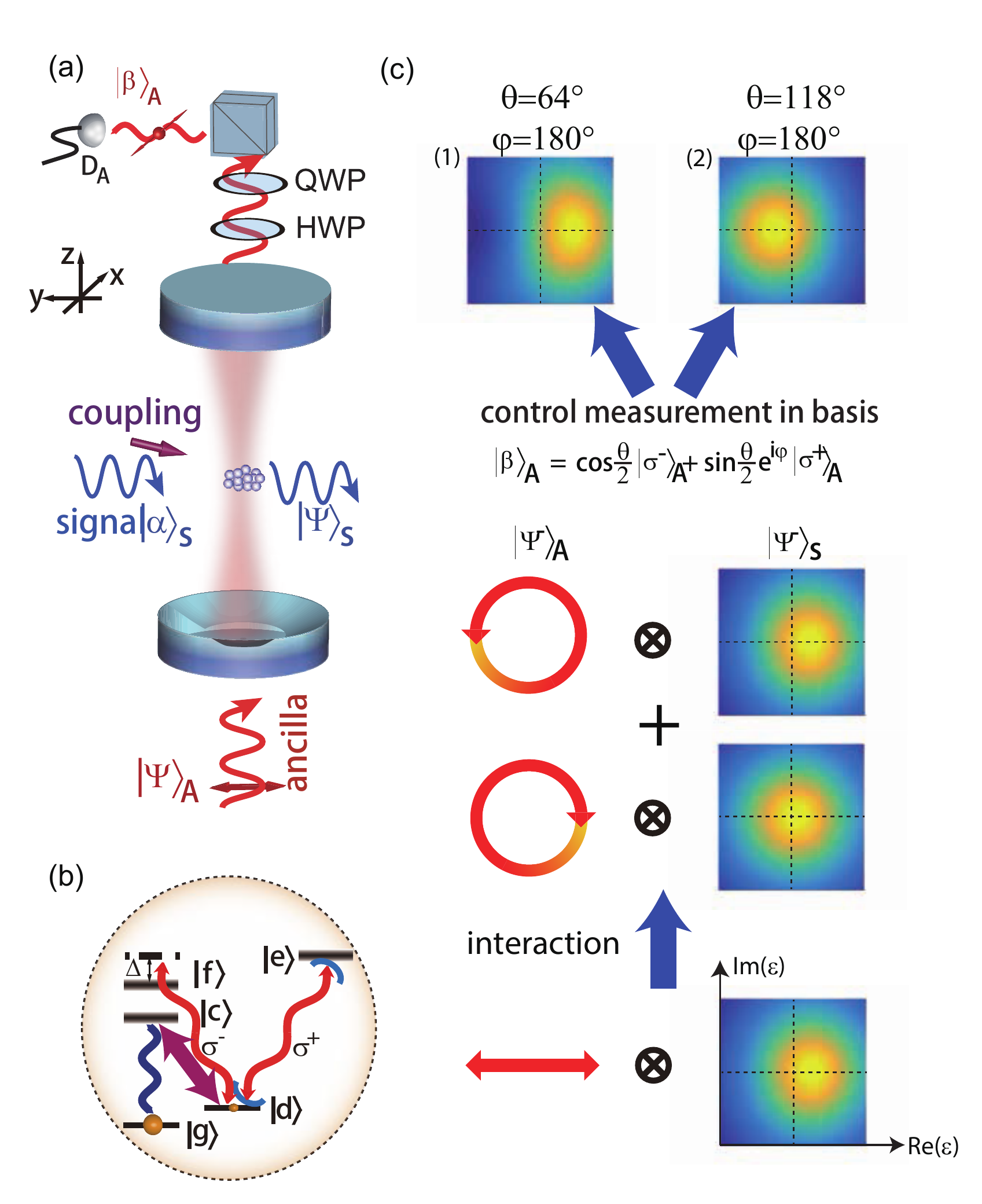}}
\caption{(a) Schematic of the experimental setup and basic idea for HIC. 
 An ensemble of cesium atoms is held in a high-finesse cavity.
 A weak signal coherent state is stored in the ensemble via EIT. For the cavity resonant with the atomic transition $\ket{d}\rightarrow\ket{e}$, weak ancilla light is sent through the cavity. The atomic excitation associated with a signal photon blocks the circularly polarized $\sigma^+$ component of the cavity light, but has little effect on the $\sigma^-$ component. The transmission of the cavity light is measured in some chosen polarization basis $\ket{\beta}_A$, decided by the angle of the half- (HWP) and quarter- (QWP)
 waveplates and the polarizing beamsplitter (PBS) preceeding the detector $D_A$. The signal state $\ket{\Psi}_S$ is then retrieved from the
 atom ensemble and measured. 
(b) Level diagram of the system. Atoms are prepared in state $\ket{g}=\ket{S_{1/2},F=3,m_F=3}$. The signal light is stored as collective excitation on $\ket{d}=\ket{S_{1/2},4,4}$ via resonant coupling to excited state $\ket{c}=\ket{P_{3/2},3,3}$. The cavity is resonant with the $\ket{d}$ to $\ket{e}=\ket{P_{3/2},5,5}$ transition.  
 (c) Changes in the ancilla mode polarization $\ket{\beta}_A$ have a large effect on the signal mode, as illustrated in (1) for $\theta=64^\circ$, $\varphi=180^\circ$ and (2) for $\theta=118^\circ$, $\varphi=180^\circ$. The plot represents a numerical calculation with mean input photon number $\aver{n_s}=0.1$. }
\label{setup}
\end{figure}

\begin{figure}[!ht]
 \includegraphics[width=1.05\columnwidth]{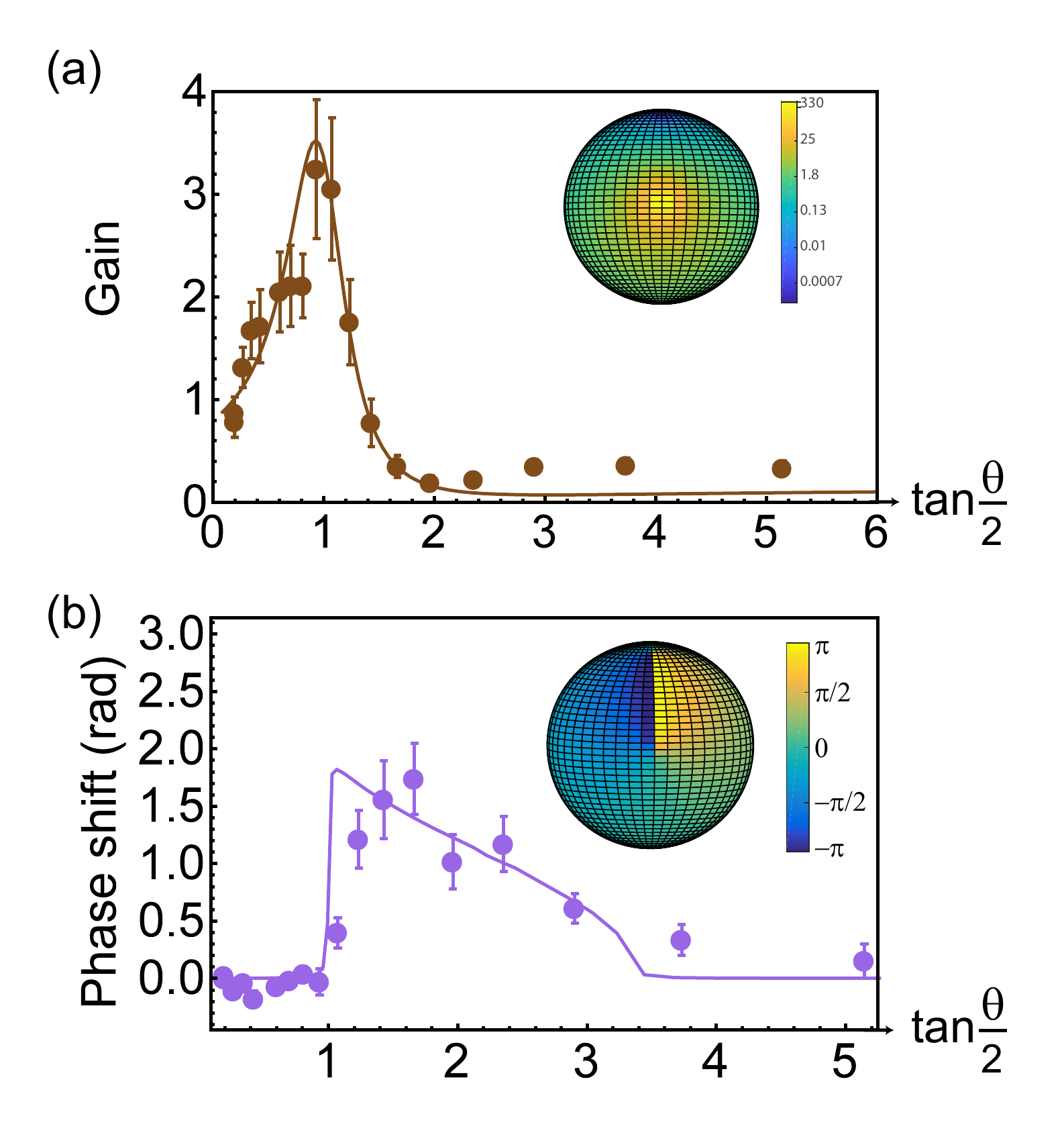}\\
\caption{Measured power gain (a) and reconstructed phase (b) of the final signal state as a function of conditioning angle $\theta$ of the ancilla mode for $\varphi=\pi$. The solid lines are predictions that include the effects of loss and non-uniform atomic coupling. Error bars in this and subsequent figures are $\pm$1 s.d. The insets show the predicted gain (a) and phase (b) as a function of the Poincar\'e sphere coordinates $\theta$ and $\varphi$ ,respectively, of the conditioning ancilla polarization for an ideal system.}
\label{phaseandamplitude}
\end{figure}

The experiment is performed with an ensemble of cold atoms in a cavity in the strong-coupling regime~\cite{Colombe:CBEC:Nature:2007,birnbaum2005photon,specht2011single,stute2012tunable}. Previously, using a similar setup, we have shown that a measurement of the ancilla mode can project the input coherent state of the signal mode into a single-photon Fock state \cite{Hosseini:PRL:2016}, and demonstrated that the phase of the signal light could be changed by about $\pi/3$ by a single ancilla photon transmitted through the cavity detuned from the atomic resonance \cite{Beck:PNAS2015}. In the current realization, we observe an anomalous and large conditional phase shift of the signal state in a near-resonant regime where the average phase shift is almost zero. The amplitude of the signal state can also be substantially changed by small changes to the conditioning polarization of the ancilla mode of a few degrees. 

In each iteration of this experiment, we use an ensemble of laser-cooled ${}^{133}$Cs atoms to create a two-mode weakly entangled state (Fig.~\ref{setup}(a) and (b)). The atoms are held inside a high-finesse cavity by a far-off-resonant dipole trap, and  prepared in the electronic ground state, $\ket{g}=\ket{S_{1/2},F=3,m_F=3}$. ($F$ and $m_F$ are the hyperfine and magnetic quantum numbers, respectively.) A weak optical coherent state with typical mean photon number $\langle n_s\rangle=0.2$ (the signal light), resonant with the $\ket{g}\rightarrow|c\rangle=|P_{3/2},3,3\rangle$ transition,  is stored in the atoms through electromagnetically induced transparency (EIT) by adiabatically reducing the power of a near-copropagating coupling laser which is resonant with the $|d\rangle=|S_{1/2},4,4\rangle\rightarrow|c\rangle$ transition. The signal-mode input coherent state $\ket{\alpha}_S$ is thus mapped onto a collective atomic excitation in the $\ket{d}$ state \cite{EIT}.  The cavity is then probed with linearly
 polarized light (ancilla light) simultaneously resonant with the cavity and the $|d\rangle\rightarrow|e\rangle=|P_{3/2},5,5\rangle$ cycling atomic transition.  (The $\sigma^-$-polarized component of the ancilla light interacts only weakly with the atoms on the $|d\rangle\rightarrow|f\rangle=|P_{3/2},5,3\rangle$
 transition.) Therefore the signal light stored in $\ket{d}$ blocks the transmission of $\sigma^+$ ancilla photons through the cavity due to the vacuum Rabi splitting \cite{vacuumRabi}, while $\sigma^-$ light is transmitted. The joint state of the light transmitted through the cavity and the retrieved signal light is a two-mode (weakly) entangled state, 
 \begin{eqnarray}
\ket{\Psi}=\ket{\sigma^-}_A \left(\ket{0}_S+\alpha\ket{1}_S\right)+\ket{\sigma^+}_A \left(\ket{0}_S+t\alpha\ket{1}_S\right)
\end{eqnarray}
where the input weak coherent signal state is approximated as $\ket{\alpha}_S\approx\ket{0}_S+\alpha\ket{1}_S$ in terms of photon Fock states, and $t$ is the transmission amplitude for $\sigma^+$-polarized light in the presence of a stored signal photon. We project the output cavity light onto a chosen polarization $\ket{\beta}_A=\cos{(\theta/2)}\ket{\sigma^-}_A+\sin{(\theta/2)}e^{i\varphi}\ket{\sigma^+}_A$, which we experimentally adjust by tuning the angles of half- (HWP) and quarter- (QWP) waveplates before the polarizing beamsplitter (PBS) in our detection path (Fig.~\ref{setup}(a)). When this projection of the ancilla photon into state $\ket{\beta}_A$ succeeds, we measure a photon click on the detector $D_A$. Simultaneously, we  measure the amplitude or  phase of the signal mode (see Supplemental Materials (SM) \cite{SM}). 
When we operate on cavity and atomic resonance, $t$ is given by $t=1/\left(1+\eta\right)$, where $\eta=8.6$ is the single-atom cooperativity~\cite{Hosseini:PRL:2016}.

Upon projection of the two-mode 
entangled state $|\Psi\rangle$ onto the polarization state 
$_A\langle{\beta}|$, the unnormalized final state $_A\bracket{\beta}{\Psi}$ is given by~\cite{martinez2017}
\begin{eqnarray}\label{sig_state_out}
&\left(\cos{\frac{\theta}{2}}+\sin{\frac{\theta}{2}}e^{i\varphi}\right)|0\rangle_S+\alpha\left(\cos{\frac{\theta}{2}} + \sin{\frac{\theta}{2}}e^{i\varphi}t\right)|1\rangle_S \nonumber \\
&\propto |0\rangle_S+\alpha \frac{\cos{(\theta/2)}+\sin{(\theta/2)}e^{i\varphi}t}{\cos{(\theta/2)}+\sin{(\theta/2)}e^{i\varphi}}|1\rangle_S = \ket{0}+\alpha' \ket{1}_S.
\end{eqnarray}
We see that depending on the ancilla detection basis, as determined by the angles $\theta$ and $\varphi$ on the Poincar\'e sphere describing ancilla light polarization, a weak coherent state $\ket{\alpha}_S$ is transformed into $\ket{\alpha'}_S$ with $\alpha'=\alpha \frac{\cos{(\theta/2)}+\sin{(\theta/2)}e^{i\varphi}t}{\cos{(\theta/2)}+\sin{(\theta/2)}e^{i\varphi}}$. The power gain of the projected coherent state is then $G=|\alpha'/\alpha|^2$. 
Recalling that transmission amplitude $t\in(0,1)$ (as determined by the interaction strength, i.e. the single atom cooperativity), and $\theta$ and  $\varphi$ are angles chosen by the measurement basis, we see that the amplitude and phase of the projected coherent state can take on any value. If we project the ancilla photon's polarization onto $\ket{\sigma^+}_A$ ($\theta=0$), the signal coherent state is unchanged. If instead we project the ancilla mode onto vertical polarization ($\theta=\pi/2,\varphi=\pi$), the signal state is maximally amplified, with the amplification attainable in the experiment set by a combination of signal-to-noise ratio and the higher-photon-number components that we have ignored in Eq.~\ref{sig_state_out}. In addition to modifying the amplitude, the choice of $\theta$ modifies the phase of the coherent state, changing it by up to $\pi$. In particular, when $\varphi=\pi$ and $t<\tan(\theta/2)<1$, the phase of the projected signal state is changed by $\pi$. In the absence of technical noise sources, this method can prepare a photonic state with strongly modified amplitude and arbitrary phase.

The measured projected phase of the signal state is shown as a function of the conditioning angle $\theta$ in Fig.~\ref{phaseandamplitude}(a). In our experiment, the maximum observed phase is limited to $\pi/2$ due to inhomogeneous coupling of atoms to the cavity light as well as dark counts of the detector. In the low photon limit, the gain of the projected signal state approximates the cross-correlation function, $g^{(2)}$ (see SM \cite{SM}), between the signal path and the cavity projection port shown in Fig.~\ref{phaseandamplitude}(b). Its maximal value is limited by background counts, which in turn limits the maximum gain in our system to $G=3.2$. To account for the inhomogeneous coupling of atoms to the cavity light we model the spatial distribution of the atoms (see SM \cite{SM}). This model takes into account the fact that our atomic cloud extends beyond the cavity mode's Gaussian waist and that atoms are randomly distributed between the nodes and antinodes of the cavity standing wave. These imperfections reduce the purity of the initial weakly entangled state $\ket{\Psi}$, and limit both the phase and gain observed in the experiment.  Moreover, background counts tend to decrease both the reconstructed phase and measured magnitude of the state (solid lines in Fig.~\ref{phaseandamplitude}(a) and (b)). When these experimental imperfections are included in the theoretical description, the model agrees well with the experimental data. 

Although we can prepare photonic states with different amplitudes and phases, we do not prepare every such state with equal probability, and states corresponding to a large change of the signal mode, or a large associated ancilla-probe interaction, are prepared more rarely. If we normalize the states $_A\langle\beta|$ and $|\Psi\rangle$, the preparation probability is simply the magnitude $|_A\langle\beta|\Psi\rangle|^2$, or the probability of observing a conditioning event before path and detector efficiency losses. 
Fig.~\ref{phaseandamplitude}(c) shows the success probability of producing signal states with fixed phase ($\varphi=\pi, \theta<-\pi/2$) as a function of the photon number gain.

\begin{figure}
 \includegraphics[width=1.05\columnwidth]{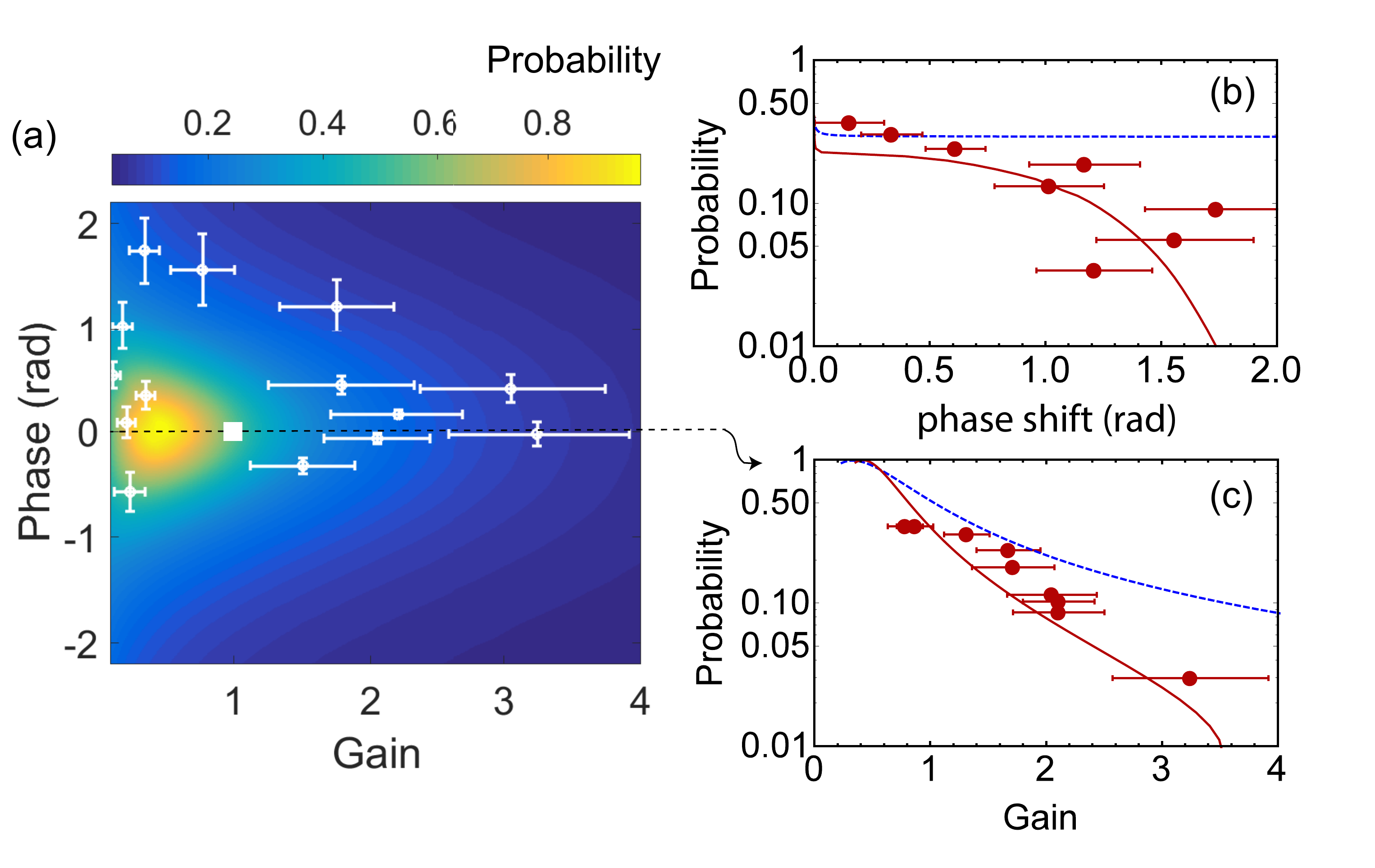}
\caption{Probability of a given interaction and signal modification. (a) The maximum theoretical probability for signal gain $G$ and phase shift $\phi_s$. The input state is indicated as a white square. A selection of the output states we produced are overlaid (circles); in particular, we produce states with both positive and negative phase shifts $\phi_s$, and both gain $G>1$ and attenuation $G<1$. This probability of state creation in our experiment is shown for (b) constant amplification (where $\theta>\pi/2, \varphi=\pi$) as a function of the conditional phase shift $\phi_s$, and (c) constant phase shift ( where $\theta<\pi/2,\varphi=\pi$) as a function of the gain $G$ of the signal state obtained from the measured correlation function, $g^{(2)}$ (see text). The solid lines in (b) and (c) represent a theoretical model taking into account the experimental imperfections. The blue dashed lines are predictions for an ideal system (i.e. one without background counts or inhomogeneous coupling) for an input state with $\langle n_s\rangle=0.2$.}
\label{prob}
\end{figure} 
 
The ideal probability of projecting into a signal state with coherent amplitude $\alpha'$ is plotted in Fig.~\ref{prob} (a) as a function of phase and relative amplitude of the final coherent state $|\alpha'\rangle_S$. The state preparation probability decreases as the projected state is displaced further from the original state (shown with a square symbol). Several experimentally projected states are shown in this figure to illustrate that we are able to produce states with different amplitudes and phases. Fig.~\ref{prob} (b) and (c) represent experimental success probability along with the theoretical prediction for a noiseless system (dashed line), and our system with experimental imperfections (solid line), as a function of phase and gain of the final signal state, respectively. A gain $G=3.2$ is achieved with a success probability of 3\%. Due to limited quantum efficiency of 0.3 of the detector, for an input signal state with a mean photon number $\langle n_s\rangle=0.2$, the detected amplified state is still within the weak coherent state limit. Thus the main deviation of the data from the theory is from higher-order excitations of the atom ensemble caused by the signal light. Provided weak enough input signal state ($G\langle n_s\rangle<1$), by confining atoms in the antinodes of the cavity standing wave and minimizing the background counts, it should be possible to achieve a conditional phase shift of $\pi$ and gain of 40 with success probabilities of $25\%$ and $1\%$, respectively.

In summary, we have demonstrated HIC for modes of light: the coherent transformation of photonic states by measurements on an ancilla mode that had previously been weakly entangled with the signal mode.  The demonstrated scheme provides a powerful tool to engineer quantum states of light by, in principle, arbitrary manipulation of their phase and amplitude. Such coherent transformation of optical states has potentially important applications in quantum communication, computations and sensing.  For example, the scheme can be used for remote state preparation \cite{Bennett:PRL2002,Kwiat:PRL2005} in quantum communication, which relies on entanglement preparation of a distant qubit conditional on the measurement outcome of another qubit without the need for Bell state measurement. The coherent amplification of optical coherent states observed here may be used to develop an optimum non-deterministic noiseless amplification \cite{Pandey2013,Ralph:PRA2014} for applications in quantum key distribution \cite{scarani2009}, state discrimination \cite{Zavatta:nphot2011}, and entanglement distillation \cite{kwiat2001,takahashi2010}.
The anomalous phase shift observed on atomic resonance can also be explained in terms of weak-value measurements \cite{Aharonov:PRL1988, Pryde:PRL2005} that has found applications in metrology \cite{Dixon:PRL2009} and understanding fundamental concepts in quantum mechanics \cite{Orozco:PRL2000, Steinberg:PRL2009}. The scheme can be also generalized to systems of massive particles and spin systems \cite{haas2014,mcconnell2015entanglement,frowis2017,zarkeshian2017}. Finally, this experiment illustrates a general paradigm that enables the heralded transformation of a quantum state that could otherwise only be accomplished by strong interactions.

\begin{acknowledgments}
KMB thanks Julian Martinez for insightful discussions. This work was supported by the NSF, the NSF CUA, and a MURI grant through AFOSR.  
\end{acknowledgments}

\bibliography{Ref}
\clearpage

\makeatletter \renewcommand{\thefigure}{S\@arabic\c@figure} \renewcommand{\thetable}{S\@arabic\c@table}  \makeatother
\makeatletter
\setcounter{figure}{0}
\makeatother
\normalsize
\onecolumngrid
\section*{Supplemental Material}
\subsection*{Apparatus} 
The Cs atoms in our experiment are held in a far off-resonant dipole trap that is focused at the cavity waist. The trap is formed by 32~mW of 937~nm  light focused through an in-vacuum lens to give an estimated transverse waist of 7 $\mu$m at the cavity mode. The corresponding approximate trap frequencies are $\omega_{\text{radial}}/(2\pi)=6$~kHz and $\omega_{\text{axial}}/(2\pi)= 0.2$~kHz. We estimate the atoms have a radial rms radius $\sigma_{radial} = 2~\mu$m and an axial rms radius $\sigma_{axial} = 50~\mu$m. The latter is perpendicular to the cavity mode with waist $w_c=35.5(2)$ $\mu$m.

The atom-cavity coupling $g$, and thus the cooperativity $\eta=4g^2/\kappa\Gamma$, varies along the standing wave of the cavity axis and with the radial extent of the cavity mode. The position and size of the atomic cloud determine the cooperativity we realize in the experiment. The maximum cooperativity $\eta_0=\frac{24\mathcal{F}/\pi}{k^2w_c^2}=8.6(1)$ is determined by the wavevector $k=2\pi/\lambda$ where $\lambda=852$~nm, the cavity waist $w_c=35.5(2)$ $\mu$m, and the cavity finesse $\mathcal{F}=77.1(5)\times10^3$. 

We model a probability distribution of the cooperativity based on the spatial distribution of the atoms. Due to pointing fluctuations of the dipole trap, along the $\hat{z}$ direction (See Fig. 2 for the coordinate system), the cooperativity varies between the
maximum value of $\eta_0$ at the antinodes of the cavity standing wave and the minimum value of 0 at the nodes. For the analysis, we average over this direction to get an effective cooperativity of $\eta_0/2$. Since the dipole trap beam waist is small compared to the cavity waist, the inhomogeneous coupling effect of the atomic cloud along the $\hat{x}$ direction is negligible. The probability of a single trial to have a cooperativity  $\eta (y) = \eta_0/2\exp{\big(-2y^2/w_c^2\big)}$ is then determined by the spatial distribution of the cloud along the $\hat{y}$ direction $p(\eta(y))=\exp{\big(-y^2/2\sigma_{axial}^2\big)}$.

\subsection*{Projected gain and probability of success} 
As we have shown in main text, we project the entangled state $\ket{\Psi}$ onto some chosen polarization of the cavity ancilla photon $\ket{\beta}_A=\cos{(\theta/2)}\ket{\sigma^-}_A+\sin{(\theta/2)}e^{i\varphi}\ket{\sigma^+}_A$. The Poincar\'e sphere coordinates $\theta$ and $\varphi$ are decided by the angles of half-(HWP) and quarter-(QWP) waveplates after the cavity, $\theta_h$, $\theta_q$,
\begin{eqnarray}
\tan{\frac{\theta}{2}}e^{i\varphi}=\frac{-i+e^{i 2 \theta_q}}{e^{4 \theta_h}-e^{i 2 (2\theta_h-\theta_q)}}.
\end{eqnarray}
In the main text, we consider the cut where $\varphi=\pi$. This corresponds to the condition of $2\theta_h=\theta_q$.

The gain introduced to the signal coherent state after projection is $G=|\alpha'/\alpha|^2=\braket{n_S|_{n_A=1}}/\braket{n_S|_{n_A=0}}$, where notation $\braket{n_a|_{n_b=k}}$ stands for the average photon counts of mode $a$ conditioned on having $k$ photon detected in mode $b$. We show here that to the lowest order, this equals to the cross correlation function $g^{(2)}_{SA}$ measured between signal and ancilla fields, 
\begin{equation} 
\begin{split}
g_{SA}^{(2)}&=\frac{\braket{n_S|_{n_A=1}}\braket{n_A}}{\braket{n_S}\braket{n_A}}=\frac{\braket{n_S|_{n_A=1}}}{\braket{n_S}}\\
&=\frac{\braket{n_S|_{n_A=1}}}{p({n_A=0})\braket{n_S|_{n_A=0}}+p({n_A=1})\braket{n_S|_{n_A=1}}}\\
&=\frac{G}{p({n_A=0})+p({n_A=1})G}
\end{split}.\label{g2}
\end{equation}
where $p({n_A=1})$ stands for the success probability of detecting a single ancilla photon in the chosen polarization, while $p({n_A=0})$ is the probability that no photon is detected. In the limit of weak coherent ancilla state, we have  $p({n_A=0})+p({n_A=1})=1$. 

From Eq. \ref{g2} we get the expression for the gain $G$:

\begin{eqnarray}
G=\frac{p({n_A=0}) g_{SA}^{(2)}}{1-p({n_A=1}) g_{SA}^{(2)}}
\end{eqnarray}

In this experiment, we tune the parameters so that the success probability is usually low (Fig.~\ref{prob}). Thus we have $p({n_A=0})\approx1$ and $p({n_A=1}) g_{SA}^{(2)}\ll 1$. In this lowest order approximation, $G\approx g^{(2)}_{SA}$. 

Considering the effect of background counts due to the polarization impurity of the ancilla photons, the measured gain $G_{exp}$ is given by
\begin{eqnarray}
G_{exp}=\frac{g_{SA}^{(2)}\times SNR+1}{SNR+1}=\frac{G\times SNR+1}{SNR+1}
\end{eqnarray}
where $SNR$ is the signal-to-noise ratio on the ancilla path that varies with projection basis and is given by (ignoring detector dark counts) $SNR=p({n_A=1})/\epsilon p({n_A=0})$. Here $\epsilon\simeq2\%$ is the measured polarization purity limited by atomic birefregence induced by run-by-run atom number fluctuation. Due to this technical imperfection, the measure gain will be averaged towards 1 (no amplification).

To calculate the success probability of projecting onto the target signal state we measure the average photon number in the ancilla port and normalize it to the total ancilla photon number exiting both ports of the PBS after the cavity.

\subsection*{Signal phase reconstruction} 
\begin{figure}
 \includegraphics[height=200pt]{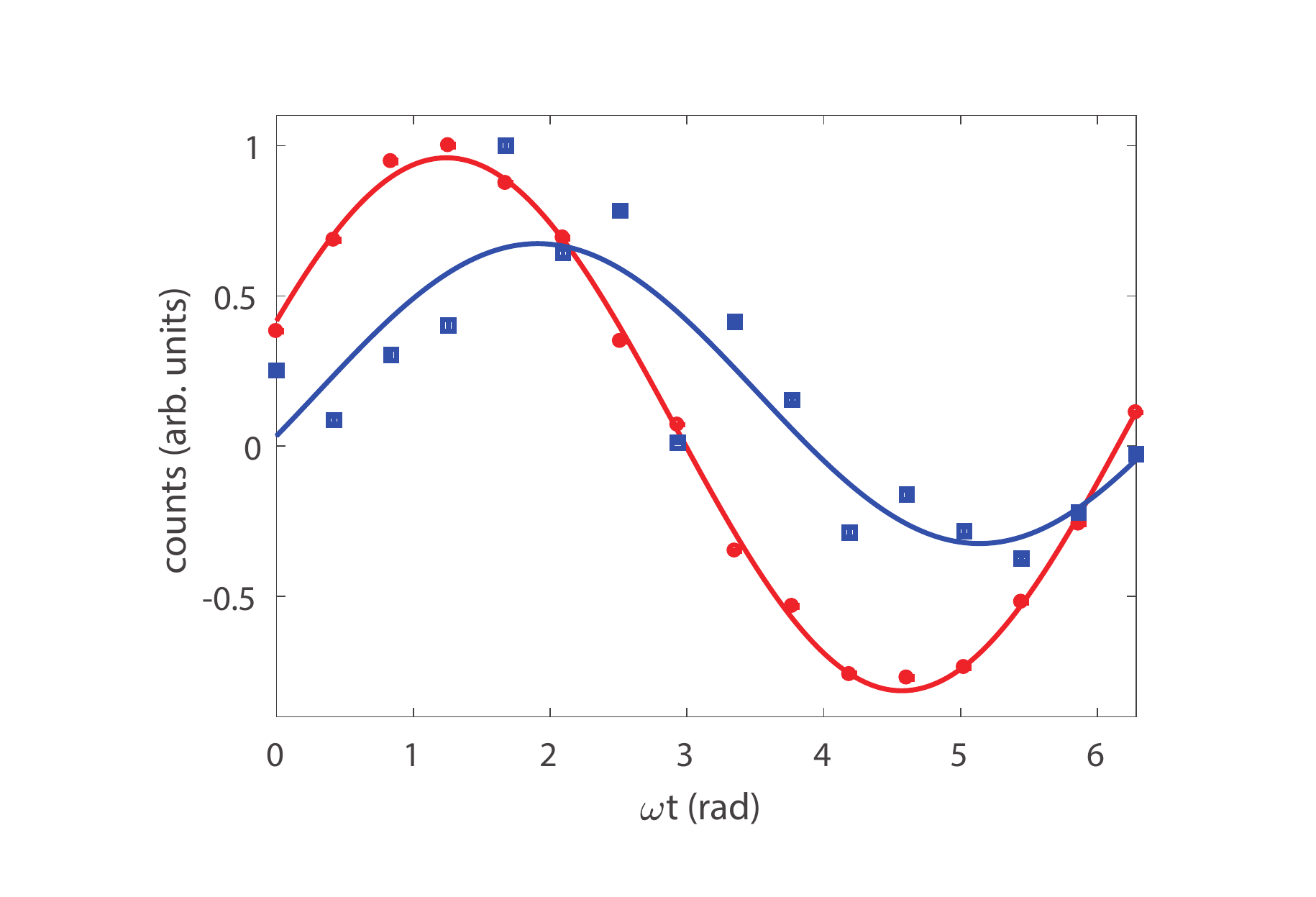}
\caption{ {\bf Signal phase reconstruction} The signal coherent state is mixed with a 30-MHz-detuned phase reference. By extracting the phase of the beatnote, we reconstruct the phase of the signal state. The red circles and blue squares and the corresponding fitting curves represent the sinusoidal beatnote for the signal state before and after the conditional interaction, respectively.}
\label{phase}
\end{figure}

To measure the phase of the signal field $\theta=\text{Arg}(\alpha')$, we mix the signal photons with a phase reference pulse which is detuned by 30 MHz from the atomic resonance. By measuring the phase of the sinusoidal beat note between them, we reconstruct the phase of the outgoing signal coherent state (Fig.~\ref{phase}).

 There are technical imperfections that limit the observation of the phase experimentally. Below we list those technical limitations and explain how their effect is included in the model.
 \begin{itemize}
 \item {\it Inhomogeneous coupling of the cavity light to the atoms.} As the atomic ensemble extends beyond the cavity mode and the atoms are distributed between nodes and anti-nodes of the cavity, different atoms couple to the cavity with different strengths. As discussed, we model this by calculating the averaged single atom cooperativity. However, when it comes to the signal phase, the varying coupling strength between the atom and the cavity also leads to a variation of the retrieved photon phase, and hence to a reduction in the contrast of the beatnote. The larger the imprinted phase is, the larger this effect will be.   
  \item {\it Atom number fluctuation and atom loss during measurement.} Each storage-interaction-retrieval experiment takes 6 $\mu$s that we repeat for 30 ms before we drop, reload the MOT and repeat the experiment cycle. During this 30ms period, due to limited life time of the atoms in the trap, the optical density decreases linearly. This shifts the cavity resonance frequency, as well as changes the birefringence induced by the atoms onto the ancilla light. These imperfections appear as an effective change in the waveplate angle over 30 ms. Similarly, atom number fluctuation due to the loading noise, can be modeled as a random variation of the waveplates' angle during measurement.  These effects can be accounted for by averaging fringes of different phase resulting from a small variation in the projection angle. These effects have not been included in the model shown in Fig.~3.
 \item {\it Background detection counts.} The signal-to-noise ratio on the ancilla detection path varies with the projection angle. Projection to a polarization with Poincar\'e sphere coordinates $\theta=\pi/2$, $\varphi=\pi$ corresponds to the lowest ancilla detection rate. When background counts are comparable to the true detection events, the phase of the projected state is mixed due to the false background events and can be estimated as:
 
  \begin{eqnarray}
\phi =\text{Arg}( P_{bg} +(1-P_{bg})e^{i \theta})
\end{eqnarray}

where $P_{bg}=\frac{\epsilon p({n_A=0})}{\epsilon p({n_A=0})+p({n_A=1})}$ is the normalized probability of having a background click. 

We note that ancilla-induced loss does not change the phase but reduces the amplitude of the signal interference fringe.

 \end{itemize}

\end{document}